\documentclass{iaus}
\usepackage{graphicx}

\title[The IACOB spectroscopic database] 
{The IACOB spectroscopic database of Galactic OB stars}

\author[S. Sim\'on-D\'iaz et al.]   
{S. Sim\'on-D\'iaz$^{1,2}$,
N. Castro$^{1,2}$,
M. Garcia$^{1,2}$
\and A. Herrero$^{1,2}$
}

\affiliation{$^1$Instituto de Astrof\'\i sica de Canarias, E-38200 
           La Laguna, Tenerife, Spain \\ 
$^2$Departamento de Astrof\'isica, Universidad de La Laguna, E-38205 
           La Laguna, Tenerife, Spain.}

\pubyear{2010}
\volume{272}  
\pagerange{119--126}
\jname{Active OB stars: structure, evolution, mass loss and critical limits}
\editors{C. Neiner, G. Wade, G. Meynet \& G. Peters, eds.}
\begin{document}

\maketitle

\begin{abstract}
We present the IACOB spectroscopic database, the largest homogeneous
database of high-resolution, high signal-to-noise ratio spectra of Northern
Galactic OB-type stars compiled up to date. The spectra were obtained with 
the FIES spectrograph attached to the Nordic Optical Telescope. 
We briefly summarize the main characeristics and present status of the IACOB, 
first scientific results, and some future plans for its extension and 
scientific exploitation.
\keywords{Stars: early-type -- Techniques: spectroscopic -- Catalogs -- 
Astronomical data bases: miscellaneous --
}
\end{abstract}

\firstsection 
\section{Introduction}

In an epoch in which we count with a new powerful generation of stellar atmopshere codes
including all the important physics for the modelling of massive OB stars, with (clusters of)
high efficiency computers allowing the computation of large grids of stellar models in more
than reasonable computational times, and with the possibility to obtain
good quality, medium resolution spectra of hundreds O and B-type stars in 
clusters outside the Milky way in just one snapshot (see e.g. the {\em FLAMES I \& II 
Surveys of Massive Stars}, \cite[Evans et al. 2008]{Eva08}, \cite[2010]{Eva10}; see
also Lennon et al., these proceedings), the compilation of medium and high-resolution 
spectroscopic databases of OB stars in our Galaxy is becoming more and more important. 
With this idea in mind, two years ago we began to compile the IACOB spectroscopic 
database, aiming at constructing the largest database
of multiepoch, high resolution, high signal-to-noise ratio (S/N) spectra of Galactic 
Northern OB-type stars. The IACOB perfectly complements the efforts also devoted in 
the last years by the GOSSS (P.I. Ma\'iz-Apellaniz; see also \cite[Sota et al., these proceedings]{Sot10})
and the OWN (P.I's Barb\'a \& Gamen, leading a multi-epoch, high-resolution spectroscopic survey
of Galactic O and WR stars in the Southern hemisphere; see \cite[Barb\'a et al. 2010]{Bar10}) teams.  

\section{Characteristics of the IACOB and present status}

We are using the FIES spectrograph\footnote{Detailed information about the NOT and FIES can be found
in http://www.not.iac.es} at the 2.56\,m Nordic Optical Telescope (NOT) in the Roque de 
los Muchachos observatory (La Palma, Spain) to compile spectra for the IACOB. A summary of the 
instrumental configuration and observing dates (before Sept. 2010) is presented in Table \ref{tab1}. 
Spectra of $\sim$100 stars with spectral types earlier than B2 and luminosity classes ranging 
from I (Supergiants) to V (Dwarfs) have already been compiled. The O-type targets were selected 
among those stars with V\,$\le$\,8 included in the GOS catalogue 
(\cite[GOSC, Ma\'iz-Apell\'aniz et al. 2004]{Mai04}). The main part of the B-type stars sample correspond 
to the works presented in \cite{Sim10a} and \cite{Sim10b}. The final spectra normally 
have S/N~$\ge$~200.

\section{Some scientific results using spectra from the IACOB}\label{section3}

There are already two published papers using data from the IACOB (and several more
in preparation). In \cite{Sim10a}, we used the stellar atmosphere code FASTWIND 
(\cite[Puls et al. 2005]{Pul05}) to perform a thorough self-consistent spectroscopic analysis of 
13 early B-type stars from the various subgroups comprising the Orion\,OB1 association; 
this study showed that the dispersion of O and Si abundances between stars in 
the various subgroups found in previous analyses (e.g. \cite[Cunha \& Lambert 1992]{Cun92}) 
was a spurious result, being the consequence of a bad characterization of the abundance 
errors propagated from the uncertainties in the stellar parameter determination. In 
\cite{Sim10b} we showed the first observational evidence for a correlation between 
macroturbulent broadening and line-profile variations in OB Supergiants using
spectroscopic time\,series for a sample of 13 OB\,Sgs; this may support the hypothesis 
that macroturbulent broadening in this type of stars is likely a result of the 
collective effect of stellar pulsations. A subsample of IACOB spectra has been used
within the {\em FLAMES-II Survey of Massive Stars: the Tarantula Survey} consortium to 
construct an atlas of medium resolution spectra of Galactic OB-type stars 
(\cite[Sana et al. in prep.]{San10}). We plan to use this atlas for the spectral 
classification of the massive stars in 30 Dor. Finally, the scientific exploitation
of the IACOB spectra concerning the quantitative spectroscopic analysis of the stars
has already began and a series of papers with results will be published soon (see 
more details in the paper devoted to the presentation of the IACOB database 
\cite[Sim\'on-D\'iaz et al., in prep.]{Sim10c}).

\section{Future plans for the IACOB}

In the next semesters, we will continue with the compilation of spectra for the IACOB,
observing stars with V$\le$8 in at least three epochs (more in the case of known
or newly detected binaries). Our idea is to make public the database via the Virtual 
Observatory in the next year. In the meantime, interested people can have access to 
the database under request to the author ({\tt ssimon@iac.es}). The complete list
of stars will be published in \cite[Sim\'on-D\'iaz et al., in prep.]{Sim10c}. We will acknowledge 
any observer who having obtained FIES spectra will like to add the spectra to the 
IACOB database after scientific exploitation.

\begin{table}[]
\caption{General characteristics of the IACOB v1.0 spectroscopic database.
}\label{tab1}
\centering
\begin{tabular}{ll|cc}
\hline
\multicolumn{2}{c}{Instrumental configuration}  & \multicolumn{2}{c}{Observing run \& Dates} \\
\hline
Telescope: NOT2.56\,m      & Spect. range:  3800\,-\,7000 \AA & 08\,A-D: 2008/11/05-08, & 10\,D: 2010/06/22 \\
Instrument: FIES           & Res. power:  46000                 & 09\,A-D: 2009/11/09-12, & 10\,E: 2010/07/15  \\
Mode: med-res              & Sampling:  0.03 \AA/pix            & 10\,A-C: 2010/06/05-07, & 10\,F: 2010/08/07 \\
\hline
\multicolumn{2}{c}{Spectral types: O4-B2 (I-V)} & \# stars: 105  & \# spectra: 720 \\
\hline
\end{tabular}
\end{table}


\begin{thebibliography}{}

\bibitem[Barb{\'a} et al.(2010)]{Bar10} 
{Barb{\'a}, R.~H., et al.} 2010, 
\textit{Rev. Mexicana AyA}, 38, 30 

\bibitem[Cunha \& Lambert(1992)]{Cun92} 
{Cunha, K., \& Lambert, D.~L.} 1992, 
\textit{ApJ}, 399, 586 

\bibitem[Evans et al.(2008)]{Eva08} 
{Evans, C.~J., et al.} 2008, 
\textit{The Messenger}, 131, 25 

\bibitem[Evans et al.(2010)]{Eva10} 
{Evans, C.~J., et al.} 2010, in: R. de Grijs and J. Lepine (eds.),
\textit{Star clusters: basic galactic building blocks throughout time and space}, Proc.
IAU Symposium No. 266, (CUP), p.\ 35 

\bibitem[Ma{\'{\i}}z-Apell{\'a}niz et al.(2004)]{Mai04} 
{Ma{\'{\i}}z-Apell{\'a}niz, J., Walborn, N.~R., Galu{\'e}, H.~{\'A}., \& Wei, L.~H.} 2004, \textit{ApJS}, 151, 103 

\bibitem[Puls et al.(2005)]{Pul05} 
{Puls, J., et al.} 2005, 
\textit{A\&A}, 435, 669 

\bibitem[Sim{\'o}n-D{\'{\i}}az (2010)]{Sim10a} 
{Sim{\'o}n-D{\'{\i}}az, S.} 2010, 
\textit {A\&A}, 510, 22 

\bibitem[Sim{\'o}n-D{\'{\i}}az et al. (2010)]{Sim10b} 
{Sim{\'o}n-D{\'{\i}}az, et al.} 2010, 
\textit{ApJ} (Letters), 720, L174


\end{thebibliography}
\end{document}